\documentstyle[amssymb,preprint,prb,aps]{revtex}

\begin{document}

\begin{titlepage}
\title
{Andreev reflection resonant tunneling through a precessing spin}
\author
{Xiufeng Cao, Yaoming Shi $^*$ and  Xiaolong Song}
\address{Department of Physics, Shanghai University,
Shanghai 200436, People's Republic of China}
\author
{Hao Chen }
\address{Department of Physics, Fudan University,
Shanghai 200433, People's Republic of China}
\maketitle

\begin{abstract}
We investigate Andreev reflection (AR) resonant tunneling through a precessing spin
which is coupled to a normal metallic lead and a superconducting lead. The formula of
 the AR conductance at zero temperature is obtained as a function of chemical 
potential and azimuthal angle of the spin precessing by using the nonequilibrium
Green function method. It is found that as the local spin precesses in a weak external
magnetic field at Larmor frequency $\omega _l$, the AR tunneling conductance
exhibits an oscillation at the frequency $2\omega _l$ alone. The amplitude of AR 
conductance oscillation enhances with spin-flip tunneling coupling increasing.
The study also shows that spin-orbit interaction in tunneling barriers is crucial for 
the oscillations of AR conductance. The effect of spin-flip tunneling coupling caused by
spin-orbit interaction and local spin precessing on resonant behavior of the AR
conductance are examined.
\\
\\
key word: Andreev reflection, precessing spin, spin-orbit interaction\\
PACS numberm:~73.63.-b,~75.20.Hr,~74. 45.+c.
\end{abstract}
\pagestyle{empty}
\end{titlepage}

1. Introduction

In recent years, the technique that is capable of single spin detection is
developed very quickly in theoretical and experimental regime. Manassen {\it %
et al}.\cite{1,2} carried out scanning tunneling microscopy (STM)
measurement of the tunneling current while scanning the surface of Si in the
vicinity of a local spin impurity (Fe cluster) or imperfection (oxygen
vacancy in Si-O) in an external magnetic field. Durkan and Welland\cite{3}
proformed a similar STM experiment on organic molecules. Above experiments
detected a small signal in the current power at the Larmor frequency.
Balatsky and Martin\cite{4} proposed a new mechanism for the spin-detection
technique -- electron spin precessing-STM. They found that in the presence
of a external magnetic field, the local spin precessing and the tunneling
current are modulated at the Larmor frequency, and the spin-flip scattering
between the injected unpolarized electron current and the local spin
produces the nodal structure of the spatial single profile. Zhu {\it et al.}%
\cite{5} studied the electronic quantum transport through a local spin
precessing in an external magnetic field in adiabatic condition. It is found
that when the spin is precessing very slowly at Larmor frequency $\omega _l$%
, the conductance develops the oscillation with the frequency of both $%
\omega _l$ and $2\omega _l$ components. The authors of Ref.[4,5] have
pointed out that spin-orbit interaction of the conduction electron in the
tunneling barriers can result in a spin-flip tunneling coupling between the
precessing spin, and the leads and the spin-flip tunneling is crucial for
electronic conductance oscillations versus the spin precessing. On the other
hand, there has been a growing interest in spin-dependent electronic
transport in mesoscopic ''hybrid'' systems \cite{6,7,8,9,10,11,12}. When one
of the leads is a superconductor, an important transport process--Andreev
reflection (AR) tunneling will occur, in which an incident electron picks up
another electron to form Cooper pair and enters the superconductor with a
hole reflected. Hence it is an interesting subject to study resonant AR
tunneling through a precessing spin.

In this letter, we mainly study AR tunneling current through a local
precessing spin (PS), which is weakly coupled to a normal metallic lead and
a superconducting lead. The schematic layout of the normal
metal-PS-superconductor (N-PS-S) system is depicted in Fig. 1, which is
different from that in Ref. [4,5]. We assume that the spin-orbit interaction
is confined in the barrier between the metallic lead and the spin site only.
It shows that the AR conductance oscillates with frequency of twice of
Larmor frequency. The amplitude of the conductance oscillations is dependent
on not only the spin-orbit interaction, but also the equilibrium chemical
potential of the system. We found that the spin-flip tunnelling coupling
caused by spin-orbit interaction plays a crucial role for the conductance
oscillations and it always enhances the oscillation amplitude of the AR
conductance.

2. model and formulation

We consider a precessing spin is coupled via tunnel barriers to a normal
metallic lead and a superconducting lead. The local spin precesses around
the weak external magnetic field applied along z-axis, which is set as the
spin quantization axis of the system shown in Fig. 1. The N-PS-S system
under consideration can be modeled by the Hamiltonian:

\begin{equation}
H=H_N+H_S+H_T+H_{PS}  \label{eq1}
\end{equation}
\noindent
with

\begin{equation}
H_N=\sum_{k\in (l),\sigma }\varepsilon _{k\sigma }a_{k\sigma }^{\dagger
}a_{k\sigma }  \label{eq2}
\end{equation}
\noindent

\begin{equation}
H_S=\sum_{k\in (r),\sigma }\varepsilon _{k\sigma }s_{k\sigma }^{\dagger
}s_{k\sigma }+\sum_{k\in (r)}(\Delta ^{*}s_{k\uparrow }^{\dagger
}s_{-k\downarrow }^{\dagger }+\Delta s_{k\uparrow }s_{-k\downarrow })
\label{eq3}
\end{equation}
\noindent

\begin{equation}
H_T=\sum_{k\in (l),\sigma ;\sigma ^{^{\prime }}}(T_{k\sigma ;\sigma
^{^{\prime }}}a_{k\sigma }^{\dagger }c_{\sigma ^{^{\prime
}}}+H.c.)+\sum_{k\in (r),\sigma }(T_{k\sigma ,\sigma }s_{k\sigma }^{\dagger
}c_\sigma +H.c.)  \label{eq4}
\end{equation}
\noindent
and

\begin{equation}
H_{PS}=J(\cos \theta c_{\uparrow }^{\dagger }c_{\uparrow }-\cos \theta
c_{\downarrow }^{\dagger }c_{\downarrow }+\sin \theta e^{-i\phi }c_{\uparrow
}^{\dagger }c_{\downarrow }+\sin \theta e^{i\phi }c_{\downarrow }^{\dagger
}c_{\uparrow })  \label{eq5}
\end{equation}
\noindent
where $H_N$ and $H_S$ are the Hamiltonians for the normal metallic lead and
the superconducting lead respectively. Under mean-field approximation, $%
\Delta $ is energy gap of the superconducting lead. $H_T$ describes the
tunneling part between the spin site and two leads with $T_{k\sigma ,\sigma
^{^{\prime }}}$ denoting the tunneling matrix. Spin-orbit interaction, which
may cause the spin-flip scattering, is considered in the barrier of the
metal side. The single electron in the region of the site is coupled to
local spin through a direction spin-exchange interaction $-g\vec{\sigma}%
\cdot \vec{S}$. Comparing with the energy of the exchange interaction, the
Zeeman energy of the electrons on the spin site in the external magnetic
field $\vec{B}$, is very small , so it can be neglected and for simplicity
the Coulomb interaction between electrons on the spin site is ignored as
well. The motion equation of the local spin is $d\vec{\mu}/dt=\vec{\mu}%
\times \gamma \vec{B}$, in which $\vec{\mu}=\gamma \vec{S}$ with $\gamma $
the gyromagnetic ratio. In the second quantization, the Hamiltonian of the
spin-exchange interaction for electrons on the site $H_{PS}$ is written as
the form in Eq.(5), in which $J$ is the effective exchange energy, $\theta $
is the tilt angle between the local spin and the external magnetic field,
and $\phi =\phi _0-\omega _lt$ is the azimuthal angle with the Larmor
frequency $\omega _l$ and the initial azimuthal angle $\phi _0$. Since the
energy associated with the spin precession, $\hslash \omega _l\backsim $ $%
10^{-6}$ eV, is much smaller than the typical electronic energy on the order
of $1$ eV, the spin precession is very slow as compared with the time scale
of all conduction electron processes. This fact allows us to treat the
electronic transport processes adiabatically, as if the local spin is static
for every instantaneous spin orientation\cite{5}. In this situation, the AR
tunneling processes studied is treated as a kind of time-independent
transport problems.

In the generalized $4\times 4$ Nambu representation, these Green's functions
of the site for non-interacting electrons can be solved exactly in the terms
of Dyson's equation, $G^{r,a}=g^{r,a}+g^{r,a}\sum^{r,a}G^{r,a}$, in which $%
\sum^{r,a}$ is the self-energy due to spin-dependent tunneling coupling and
the off-diagonal elements of the local spin processing $J\sin \theta e^{\pm
i\phi }$, and $g^{r,a}$ is the Green function without perturbation and
spin-flip scattering on the spin site: 
\begin{equation}
(g^{r,a})^{-1}=\left( 
\begin{array}{llll}
\omega -J\cos \theta \pm i\delta ^{+} & \hspace{0.4in}0 & \hspace{0.4in}0 & 
\hspace{0.4in}0 \\ 
\hspace{0.4in}0 & \omega +J\cos \theta \pm i\delta ^{+} & \hspace{0.4in}0 & 
\hspace{0.4in}0 \\ 
\hspace{0.4in}0 & \hspace{0.4in}0 & \omega -J\cos \theta \pm i\delta ^{+} & 
\hspace{0.4in}0 \\ 
\hspace{0.4in}0 & \hspace{0.4in}0 & \hspace{0.4in}0 & \omega +J\cos \theta
\pm i\delta ^{+}
\end{array}
\right)  \label{eq6}
\end{equation}
\noindent
For the F-PS-S system, the $\sum^{r,a}$ is written as $\sum^{r,a}=\sum_{ps}+%
\sum_n^{r,a}+\sum_s^{r,a}$. Here the off-diagonal term of $H_{PS}$ is
considered by self-energy $\Sigma _{ps}$ with:

\begin{equation}
\Sigma _{ps}=J\sin \theta \left( 
\begin{array}{llll}
0 & 0 & e^{i\phi } & 0 \\ 
0 & 0 & 0 & -e^{i\phi } \\ 
e^{-i\phi } & 0 & 0 & 0 \\ 
0 & -e^{-i\phi } & 0 & 0
\end{array}
\right)  \label{eq7}
\end{equation}
\noindent

Within the wide bandwidth approximation, the self-energy $\sum_n^{r,a}$
coupling to the normal metallic lead is evaluated from $\sum_n^{r,a}=\mp 
\frac i2\Gamma _n$ with

\begin{equation}
\Gamma _n=\Gamma _0\left( 
\begin{array}{llll}
1+\lambda ^2 & 0 & 2\lambda & 0 \\ 
0 & 1+\lambda ^2 & 0 & 2\lambda \\ 
2\lambda & 0 & 1+\lambda ^2 & 0 \\ 
0 & 2\lambda & 0 & 1+\lambda ^2
\end{array}
\right)  \label{eq8}
\end{equation}
\noindent
where $\lambda $ is defined as a ratio of the spin-flip and spin-unflip
tunneling amplitude, $\lambda =\mid T_{k\sigma ,\bar{\sigma}}\mid /\mid
T_{k\sigma ,\sigma }\mid $. $\Gamma _0=2\pi T_{k\sigma ,\sigma }^{*}\rho
_nT_{k\sigma ,\sigma }$ is the tunneling coupling without spin-flip
scattering. Thus the tunneling couplings associated with the spin-flip
tunneling amplitude, are expressed as $\lambda \Gamma _0=2\pi T_{k\sigma ,%
\bar{\sigma}}^{*}\rho _nT_{k\sigma ,\sigma }$ or $2\pi T_{k\sigma ,\sigma
}^{*}\rho _nT_{k\sigma ,\bar{\sigma}}$, and $\lambda ^2\Gamma _0=2\pi
T_{k\sigma ,\bar{\sigma}}^{*}\rho _nT_{k\sigma ,\bar{\sigma}}$, due to
spin-orbit interaction in barrier of metal side. The self-energy coupling to
the S-lead is:

\begin{equation}
\Sigma _s^{r,a}=\mp \frac i2\rho _s^r(\omega )\Gamma _0\left( 
\begin{array}{llll}
1 & -\frac \Delta \omega & 0 & 0 \\ 
-\frac \Delta \omega & 1 & 0 & 0 \\ 
0 & 0 & 1 & \frac \Delta \omega \\ 
0 & 0 & \frac \Delta \omega & 1
\end{array}
\right)  \label{eq9}
\end{equation}
\noindent
where $\rho _s^r(\omega )$ is the dimensionless BCS density of states: 
\begin{equation}
\rho _s^r(\omega )=\frac{\left| \omega \right| \theta (\left| \omega \right|
-\Delta )}{\sqrt{\omega ^2-\Delta ^2}}+\frac{\left| \omega \right| \theta
(\Delta -\left| \omega \right| )}{i\sqrt{\Delta ^2-\omega ^2}}  \label{eq10}
\end{equation}
\noindent
For convenience, we introduce the linewidth function matrix coupling to the
S-lead:

\begin{equation}
\Gamma _s=\rho _s^{<}(\omega )\Gamma _0\left( 
\begin{array}{llll}
1 & -\frac \Delta \omega & 0 & 0 \\ 
-\frac \Delta \omega & 1 & 0 & 0 \\ 
0 & 0 & 1 & \frac \Delta \omega \\ 
0 & 0 & \frac \Delta \omega & 1
\end{array}
\right)  \label{eq11}
\end{equation}
\noindent
with $\rho _s^{<}(\omega )=\left| \omega \right| \theta (\left| \omega
\right| -\Delta )/\sqrt{\omega ^2-\Delta ^2}$. After a straightforward
calculation, the normal electron tunneling conductance and the Andreev
reflection conductance are obtained in the linear response regime as
follows: 
\begin{equation}
G_N=\frac{e^2}h\int d\omega [-\frac{\partial f}{\partial \omega }%
]\sum_{i=1,3}[G^r\Gamma _sG^a\Gamma _n]_{ii}  \label{eq12}
\end{equation}
\noindent
and 
\begin{equation}
G_A=\frac{2e^2}h\int d\omega [-\frac{\partial f}{\partial \omega }%
]\sum_{i=1,3}^{j=2,4}G_{ij}^r(\Gamma _nG^a\Gamma _n)_{ji}  \label{eq13}
\end{equation}
\noindent

Since normal linear conductance is zero, $G_N=0$, at zero temperature, only
the Andreev reflection process contributes to electronic transport of the
system. So the total conductance $G$ is equivalent to $G_A$.

3. The results and discussion

We only concentrate here on the case of the spin-exchange interaction
strength $J$ is restricted in the range of energy gap of the superconductor $%
\Delta $ ( $J\leq \Delta $). In the following calculation, $\Delta $ is
taken as energy unit and the spin-exchange interaction strength is chosen as 
$J=0.5$.

In order to examine resonant behaviors of the AR conductance versus the
chemical potential $\mu $, we first consider AR tunneling processes through
a static spin in zero magnetic field. In Fig. 2, we plot the conductance
versus chemical potential $\mu $ with some different values of the ratio of
spin-dependent tunneling amplitude, $\lambda =0.0$ (solid line), $0.4$
(dashed line), $0.8$ (dotted line). The curves shown in panels (a), (b) and
(c) correspond to three different orientations of the local spin: $(\theta
,\phi )=(0,0),$ $(\pi /4,0),$ $(\pi /2,0)$, in which the tunneling coupling
without spin-unflip scattering is taken as $\Gamma _0=0.1$. There are
several generic features of resonant AR conductances in (a), (b) and (c) of
Fig.2. It is clearly seen that two resonant peaks of the conductance appear
symmetrically at the two sides of $\mu =0$, due to the Andreev reflection is
determined by spin minority population\cite{9}. Moreover the position of
every resonant peak is almost independent on the relative spin-dependent
tunneling amplitude $\lambda $, and has a small deviation from $\mu $ $=$ $%
J(-J)$. This is different from previous results in normal electron tunneling
conductance in Ref.[5]. The spin-orbit interaction in tunneling barrier
influences resonant amplitude of the AR conductance, which is different from
Ref.[5]. Comparing Fig. 2(a)-2(c), it is clearly seen that in the case of $%
\theta =0$, the spin-orbit interaction strongly suppresses the conductance
not only at $\lambda =0.4$, but also $\lambda =0.8$. In the case of $\theta
\neq 0$, however, the AR\ conductance is efficiently suppressed only for
strong spin-orbit interaction $\lambda =0.8$. For $\theta =0$, spin-flip
scattering is dominated by spin-orbit interaction, but for $\theta \neq 0,$
spin-orbit interaction is only one part of the spin-flip scattering, which
also involves the component of precessing spin, $J\sin \theta e^{\pm i\phi }$%
.

Fig. 3 presents some curves of the AR conductance oscillations of as a
function of the phase $\phi $ in units of $2\pi $, which a tunneling
electron accumulates from the precession of the local spin, with various
values of $\lambda =0.0$ (solid line), $0.4$ (dashed line), $0.8$ (dotted
line) for given parameters $\Gamma _0=0.1$, $\theta =\pi /2$, and $\phi _0=0$%
. The results shown in the left (a), (b) and (c) panels of Fig.3 correspond
to three different chemical potentials, (a) $\mu =0.1$, (b) $0.4$, and (c) $%
0.7$, respectively. The Fourier spectrums of the AR conductance oscillations
with $\lambda =0.4$ are presented in the right three corresponding panels of
Fig.3, in which other parameters are the same as in the left panels of
Fig.3. The calculated result exhibits obviously that oscillation of the AR
conductance occurs only in the case of the spin-flip tunneling coupling and
its oscillation amplitude enhances with spin-flip tunneling coupling. The
same conclusion was given in Ref. [4,5].

It is well known, in the presence of magnetic field, the local spin will
precess with the Larmor frequency $\omega _l$. The question is how will this
Larmor precession influence the conductance of electron transported through
the local spin. We find that the AR conductance will oscillate at double
Larmor frequency $2\omega _l$ alone, because the Andreev reflection
conductance is usually expressed by the off-diagonal terms of the Green
functions\cite{7}, such as $T^A=\Gamma _l^2\left| G_{12}^r(\omega )\right|
^2 $. Moreover the off-diagonal terms of Green functions often satisfy the
relation of $G_{ij}(\phi +\pi )=G_{ji}(\phi )$ \cite{5}, which results in
the change of the oscillation period from $2\pi $ to $\pi $.The occurrence
of non-oscillatory conductance is understood as follow. The spin-flip
scattering of transited electrons is determined only by the scattering terms
of the exchange interaction, $J\sin \theta e^{\pm i\phi }$, on the
spin-site, if there is no spin-orbit interaction in the tunneling barriers..
Due to the above two terms, $J\sin \theta e^{i\phi }$ and $J\sin \theta
e^{-i\phi }$, are out-phase for $\phi $, the tunneling conductance, which is
proportional to the absolute value squared of transmission amplitude, should
carry no information of the spin precessing i.e. the azimuthal angle $\phi $%
. However, in the presence of spin-orbit interaction in the tunneling
barriers, the spin-flip scattering amplitude of electrons should be
expressed as $J\sin \theta e^{\pm i\phi }+2\lambda \Gamma _n$, so the
information of $\phi $ can be contained in the multiplication of $(J\sin
\theta e^{\pm i\phi }+2\lambda \Gamma _n)$ and $(J\sin \theta e^{\pm i\phi
}+2\lambda \Gamma _n)^{*}$, and in the tunneling conductance of the system.

The oscillation amplitude increases with the spin-orbit interaction and the
oscillation frequency is twice of Larmor frequency. Comparing Fig.
3(a)-3(c), we obtain that the oscillation amplitude is modulated by the
equilibrium chemical potential except spin-orbit interaction. When chemical
potential trends towards exchange interaction strength $J$ from two sides,
the oscillation amplitude increases. These features can be seen more clearly
from the Fourier spectrum.

In summary, we have studied AR resonant tunneling through a local spin
precessing in the external magnetic field, which is coupling to normal
metallic and superconducting leads. It is found that the spin-orbit
interaction in the tunneling barriers between the spin site and metallic
lead is crucial for the appearance of AR conductance oscillations versus the
azimuthal angle of the spin precessing $\phi $. The conductance oscillation
is modulated by spin-orbit interaction and the equilibrium chemical
potential of the system. The oscillation amplitude of AR conductance
enhances with spin-flip tunneling coupling increasing. The study shows that
the AR tunneling conductance exhibits a oscillation at the frequency double
Larmor frequency $2\omega _l$ alone. The technique combining STM with
superconductor must be a new test of the proposed mechanism for the
conductance oscillation.

Acknowledgments: The authors are grateful to Sun Qing-feng for meaningful
discussion.. This work was supported by the Natural Science Foundation of
China (NSFC) under Projects No.90206031, and the National Key Program of
Basic Research Development of China (Grant No. G2000067107). It was also
supported by the National Natural Science Foundation of china (Grant No.
60371033) and Shanghai Leading Academic Discipline Program, China.

\begin{description}
\item  
\begin{center}
{\large FIGURES}
\end{center}

\vspace{0.3cm} 
\vspace{0.3cm} 

\item  {FIG.~1 Schematics of the system investigated: normal metallic (N)
and superconducting (S) lead attached to a local spin ($\vec{S}$}{), which
precesses around the magnetic field ($\vec{B}$}{) with tilt angle }$\theta $.


\vspace{0.3cm} 
\vspace{0.3cm} 

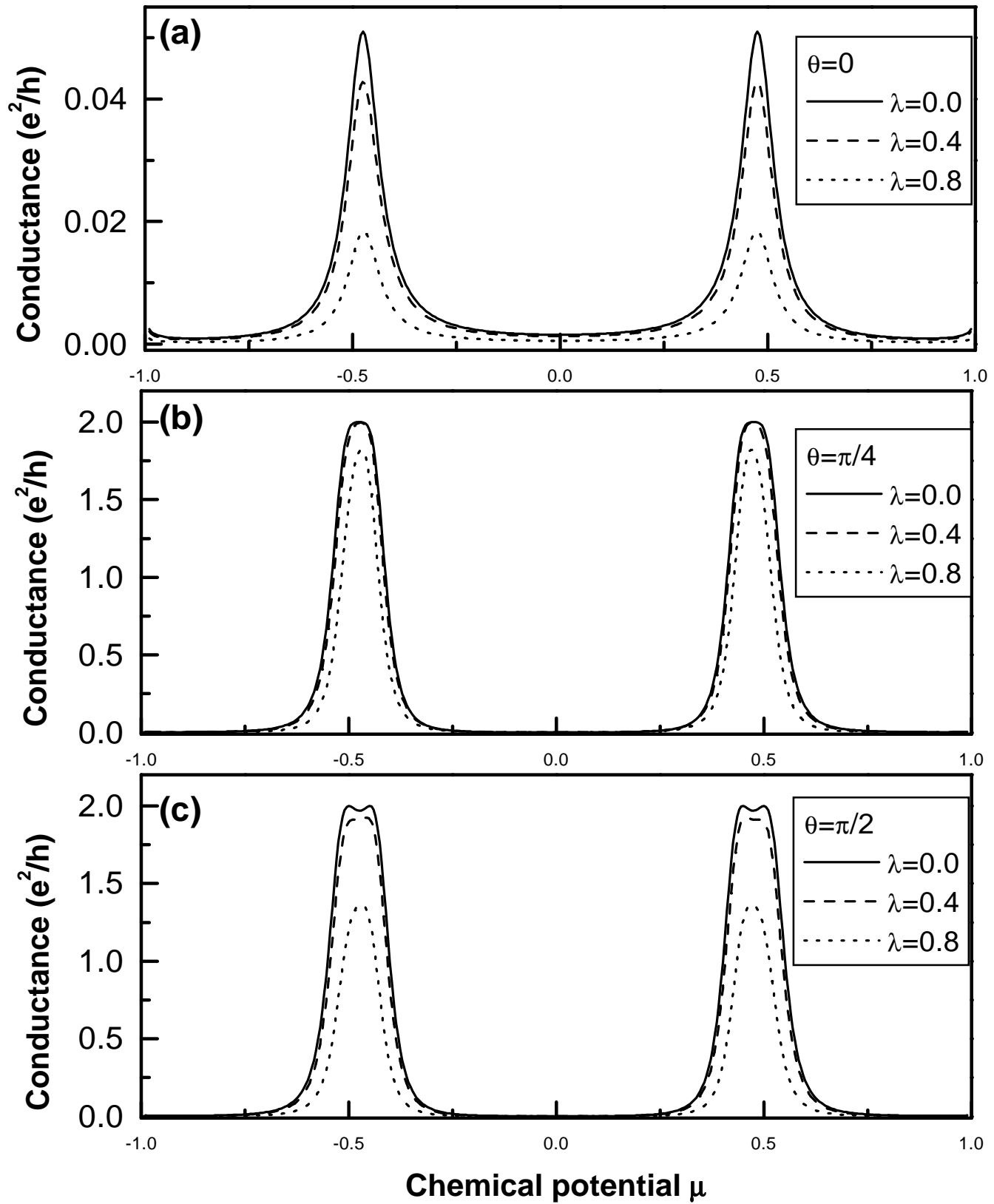
\item  {FIG.~2. The linear resonant AR conductance vs chemical potential }$%
\mu ${\ with }$\Gamma _0=0.1$ for different ratio of the spin-flip and
spin-unflip tunneling amplitudes: $\lambda =0.0$ (solid line), $0.4$ (dashed
line), $0.8$ (dotted line). The curves shown in panels (a) through (c)
correspond to three different precessing orientations: $({\theta ,\phi )}%
=(0,0)$, $(\pi /4,0)$, $(\pi /2,0)$.


\vspace{0.3cm} 
\vspace{0.3cm} 

\item  {FIG.~3 The conductance versus the phase }$\phi $ with various values
of the ratio of the spin-flip and spin-unflip tunneling amplitudes, $\lambda
=0.0$ (solid line), $0.4$ (dashed line), $0.8$ (dotted line). The curves
shown in the left panels (a) through (c) correspond to three different
values of the chemical potential: $\mu =0.1$, $0.4$, $0.7.$ Also shown with
the right panels are the Fourier spectrum for $\lambda =0.4$ with the
chemical potential same as the left panels. Other parameter values: $\Gamma
_0=0.1$, $\theta =\pi /2$ and $\phi _0=0$.

\end{description}

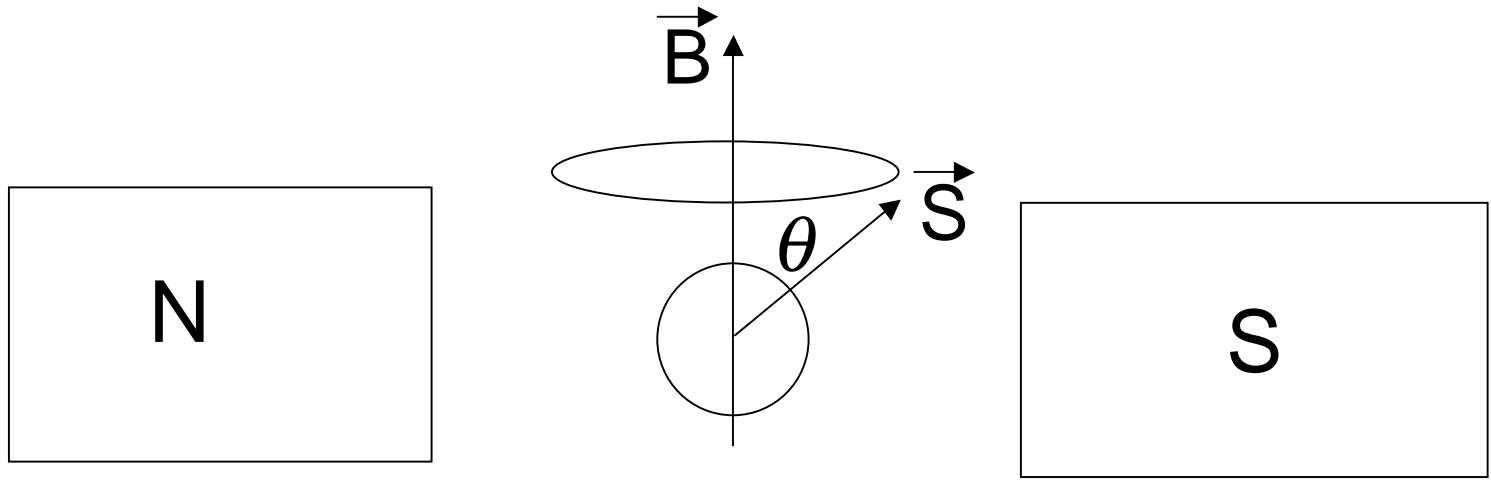
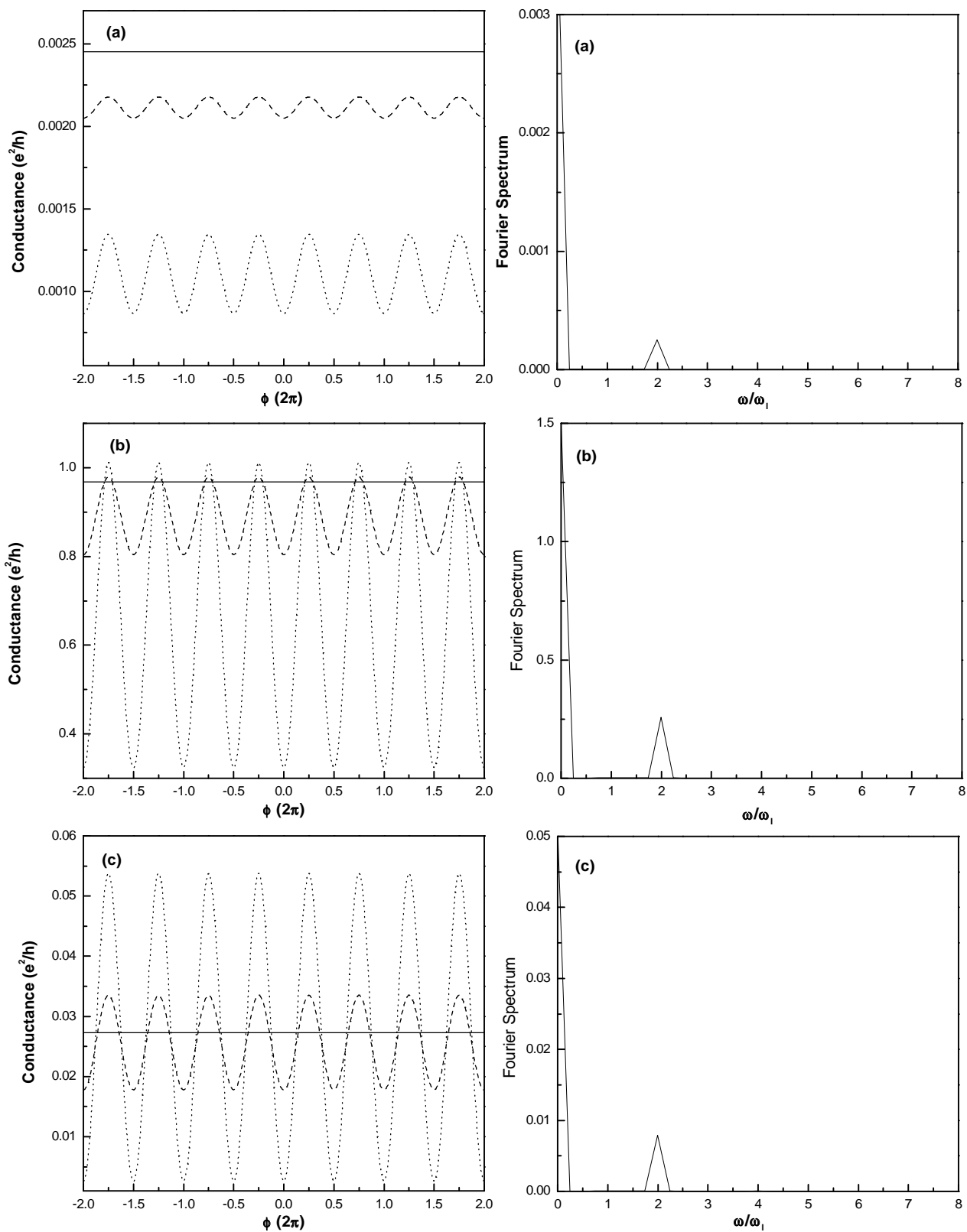
\end{document}